\documentclass[letterpaper, 10 pt, conference]{ieeeconf}  

\IEEEoverridecommandlockouts                              

\usepackage{amsmath,amssymb,amsfonts}
\usepackage{graphicx}
\usepackage{tabularx}
\usepackage{textcomp}
\usepackage{xcolor}
\usepackage{algorithmic}
\usepackage{cite} 
\usepackage{subfig}
\usepackage{url}

\title{\LARGE \bf
Developing a Modular Toolkit for Rapid Prototyping of Wearable Vibrotactile Haptic Harness
}

\author{Sandeep Kollannur$^{1}$, Katherine (Katie) Robertson$^{2}$, and Heather Culbertson$^{3}$
\thanks{*This work was supported by Innovation in Haptics fund (Meta Inc) from The IEEE TC on Haptics of the IEEE RAS. (2022)}
\thanks{$^{1}$Sandeep Kollannur is with the Department of Computer Science, University of Southern California,
        Los Angeles, USA
        {\tt\small sandeep.kollannur@usc.edu}}%
\thanks{$^{2}$Katherine (Katie) Robertson is with California Institute of Technology, Pasadena, CA 91125, USA
         {\tt\small kcrobert@caltech.edu }}%
\thanks{$^{3}$Heather Culbertson is with the Department of Computer Science, University of Southern California,
        Los Angeles, USA
        {\tt\small hculbert@usc.edu}}%
}

\begin{document}

\maketitle
\thispagestyle{empty}
\pagestyle{empty}

\begin{abstract}
This paper presents a toolkit for rapid harness prototyping. These wearable structures attach vibrotactile actuators to the body using modular elements like 3D printed joints, laser cut or vinyl cutter-based sheets and magnetic clasps. This facilitates easy customization and assembly. The toolkit's primary objective is to simplify the design of haptic wearables, making research in this field easier and more approachable.
\end{abstract}

\section{Introduction}
Several research studies in haptics rely on custom hardware, but building prototypes from scratch can be challenging\cite{b4}. Researchers must design systems to mount actuators on the body while considering factors like mobility, degrees of freedom, body proportions, ease of donning/doffing, and sanitation\cite{b5}. Similarly Wearable Bits\cite{3374920.3374954} has a scaffolding platform for wearable e-textiles. Our proposal aims to simplify the prototyping of wearable devices by providing a toolkit that uses modular, customizable components to create harness structures for mounting actuators on the body. Our work aims to simplify the  process of prototyping wearable devices by providing a toolkit that uses modular, customizable components to create harness structures for mounting vibrotactile actuators on the body. 

A harness comprises a structured framework that people wear on their body, resembling a network of straps, belts, or webbing. Users can adjust it to fit snugly while providing mounting points for various devices or tools. Constructing such harnesses requires expertise in mechanics, electronics (wiring), and ergonomics\cite{b4}. By providing a framework of easy to manufacture joints, clamps, and structural elements, this toolkit lowers the barrier for haptics researchers to quickly build prototypes tailored to their experiments\cite{b8}. Enabling easier customization also facilitates reproducibility and sharing of insights across different studies\cite{b1}. Overall, this toolkit will accelerate haptics research\cite{b3}.

\section{Design Principles}

The principles guiding the toolkit design are modularity, customizability, and accessibility\cite{b3}. Off-the-shelf components, 3D printed joints, and magnetic clasps can be easily connected and modified\cite{b1}. Parametric CAD models and open-source plans enable researchers to customize parts as needed\cite{b11}. Low-cost off-the-shelf materials make the toolkit financially accessible\cite{b4}. These principles allow rapid assembly of harnesses for diverse applications, for example, the forearm, legs, knees, neck and other body locations\cite{b5}.

The design principles of the toolkit design are based on our analysis of the literature in haptic design:

\begin{enumerate}
    \item \textbf{Degrees of freedom (DOF)\cite{b5}:} Restricting wearer's movement can impact experimental validity. The harness should aim to be low-profile and avoid hindering natural motions.
    \item \textbf{Sanitization\cite{b3}:} Components should be materials/finishes that can be thoroughly cleaned between users to prevent cross-contamination and to provide a safe experience for participants of research studies.
    \item \textbf{Robustness and durability\cite{b3}:} For consistent results, the harness must withstand wear and tear over many trials without failure.  
    \item \textbf{Ease of don/doffing\cite{b1}:} Complex harnesses that are difficult to put on and take off will slow experiment pacing and potentially frustrate users.
    \item \textbf{Flexibility and modularity\cite{b11}:} Components that can be easily substituted will maximize customizability for diverse needs.
    \item \textbf{Accessibility\cite{b4}}
        \begin{itemize}
            \item \textbf{Low-cost materials\cite{b4}:} Utilizing commodity materials lowers barriers for access and iteration.  
            \item \textbf{Off-the-shelf materials\cite{b4}:} Enables immediate manufacturing without custom fabrication. 
            \item \textbf{Ease of customization\cite{b1}:} Researchers and designers of all skill levels should be able to quickly modify and adapt the toolkit.
        \end{itemize}
    \item \textbf{User groups' perspectives\cite{b3}}
        \begin{itemize}
            \item \textbf{Haptic experience designer\cite{b4}:} Must enable the designer's creative vision and rapid iteration to test concepts. 
            \item \textbf{Experimenter\cite{b4}:} Needs to easily and quickly mount harness.
            \item \textbf{Participant\cite{b5}:} Comfort and unhindered natural movement is critical for avoiding fatigue and behavior changes. Safety and perceived quality also matter.
        \end{itemize}
\end{enumerate}

Applying these human-centric principles allows the creation of harnesses tailored for diverse applications, users, and research needs\cite{b5}. The toolkit should empower all stakeholders in the process.

\section{Design Updates and Process}
The first framework involves a series of interconnected actuators (Fig. 1a). Initially, the system used silicone tubes as connecting leads. Inserting different materials (i.e. metal) within the tubes customizes their flexibility. Magnetic disks at the openings of the silicone tubes form clasps for easy donning and doffing. The silicone tubes connect with 3D-printed joints of a defined number of spokes to form a supportive network for 3D-printed friction holders for actuators.

\captionsetup[figure]{justification=centering}

Later, the silicone tubes and joints were replaced by straps and tiles. Programmatically changing the geometry of the tiles (Fig. 2b) allows for discrete configurations (Fig. 2c). The connective elastic and nylon straps can adjust to produce the desired normal forces and positions of the actuators, which mount upon the tiles via a magnetic ring (Fig. 2a). 

\begin{figure}[tb]
    \centering
    \hspace*{\fill}
    \subfloat[Actuators layout(s)\label{fig:Tyvek tile}]{
        \includegraphics[width=0.15\columnwidth, angle = 0]{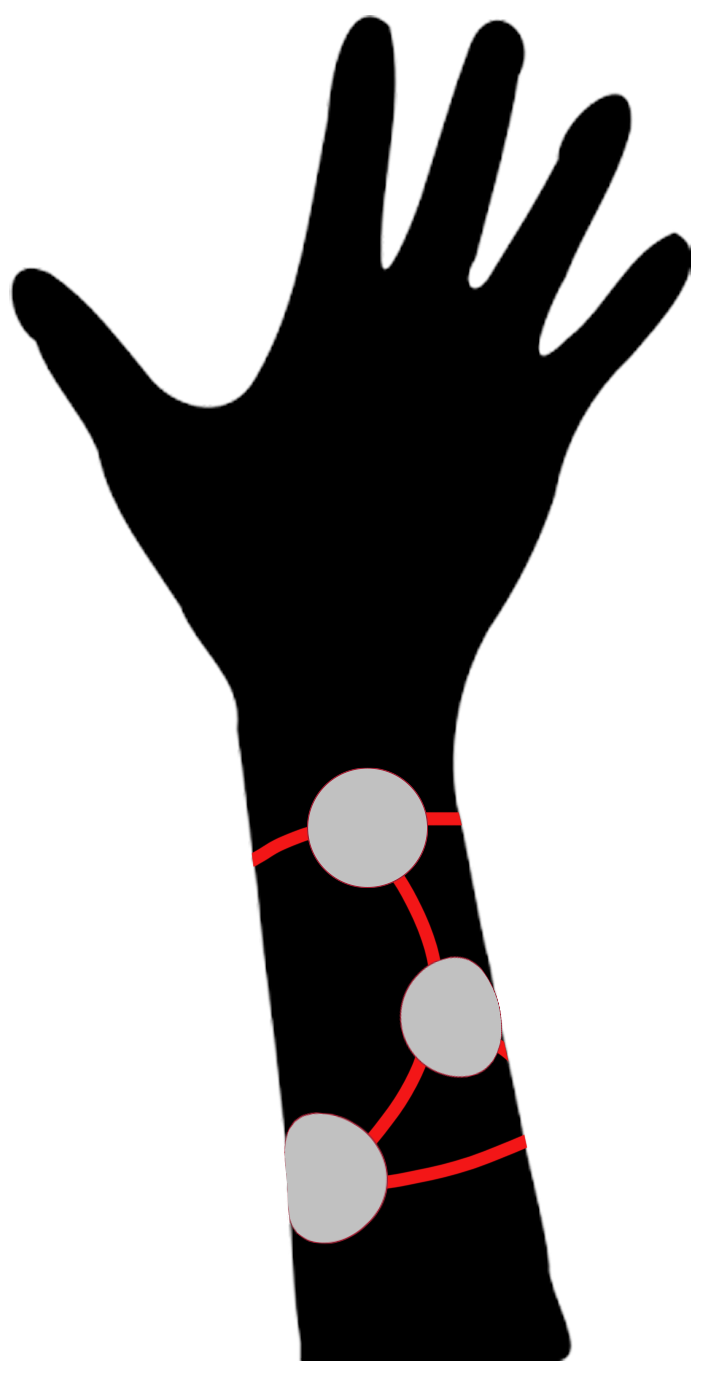}
    }
    \hfill
    \subfloat[TPU based mount\label{fig:TPU tile}]{
        \includegraphics[width=0.4\columnwidth]{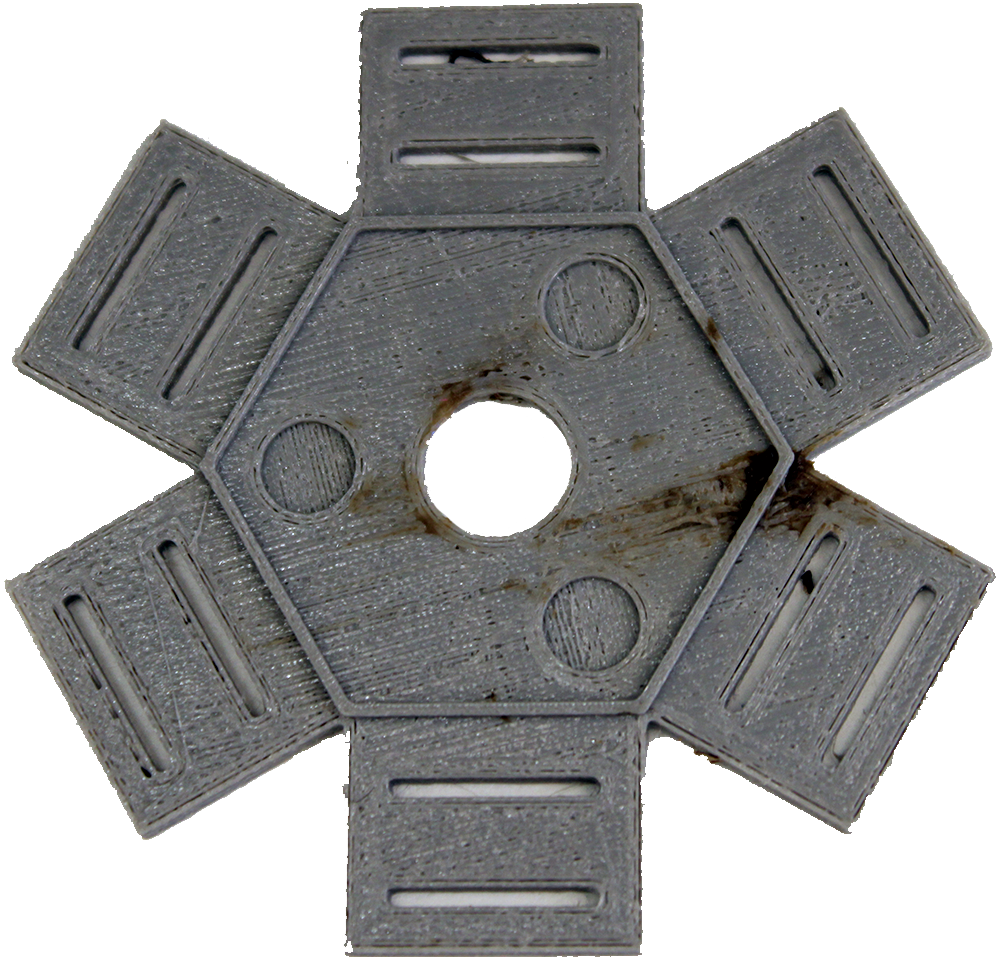}
    }
    \hspace*{\fill}
    \caption{First design scheme}
    \label{fig:example tile materials}
\end{figure}

The tile system uses a lamination of laser-cut \texttt{Tyvek}\footnote{\url{https://www.dupont.com/brands/tyvek.html}} and EVA foam to provide comfort and durability, and a 3D-printed ring of magnets to mount an actuator.

\begin{figure}[tb]
    \centering
    \begin{tabularx}{\columnwidth}{cc}
        \subfloat[Actuators mount on tile\label{front of tile}]{
            \includegraphics[width=0.4\columnwidth]{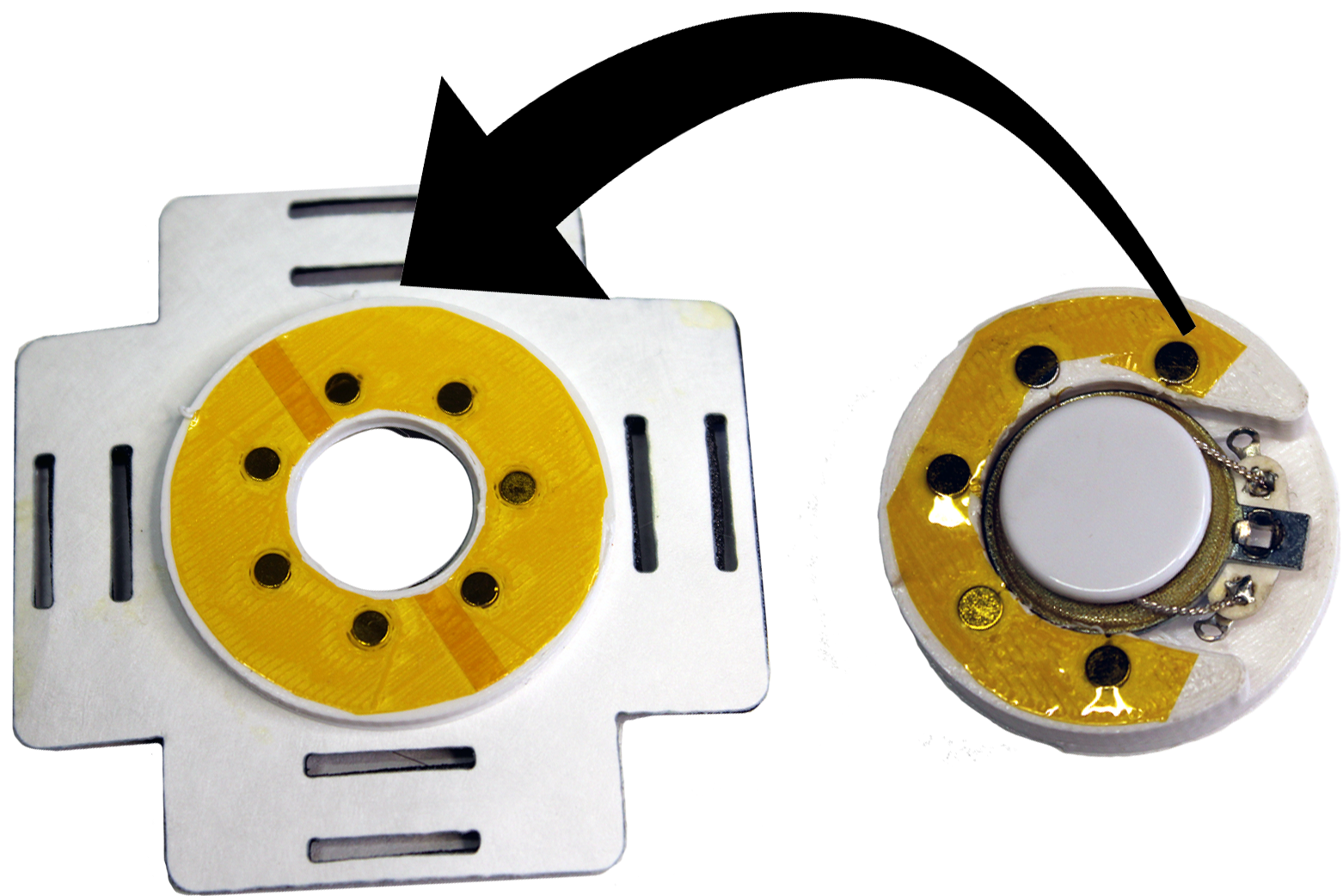}
        } &
        \subfloat[Examples of tile geometries\label{tile geometries}]{
            \includegraphics[width=0.4\columnwidth]{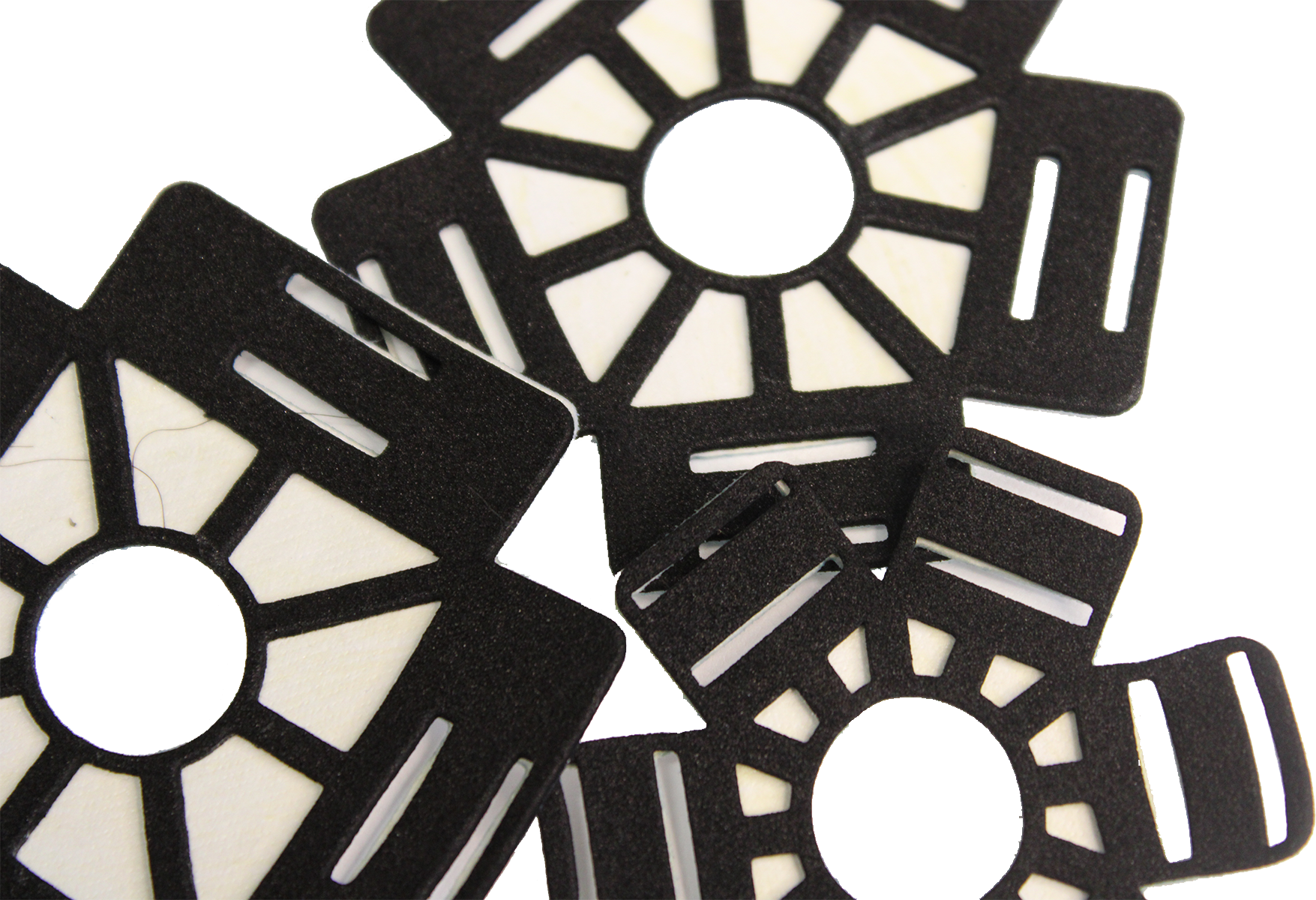}
        } \\
        \multicolumn{2}{c}{
            \subfloat[A basic tile configuration\label{basic tile configuration}]{
                \includegraphics[width=0.4\columnwidth]{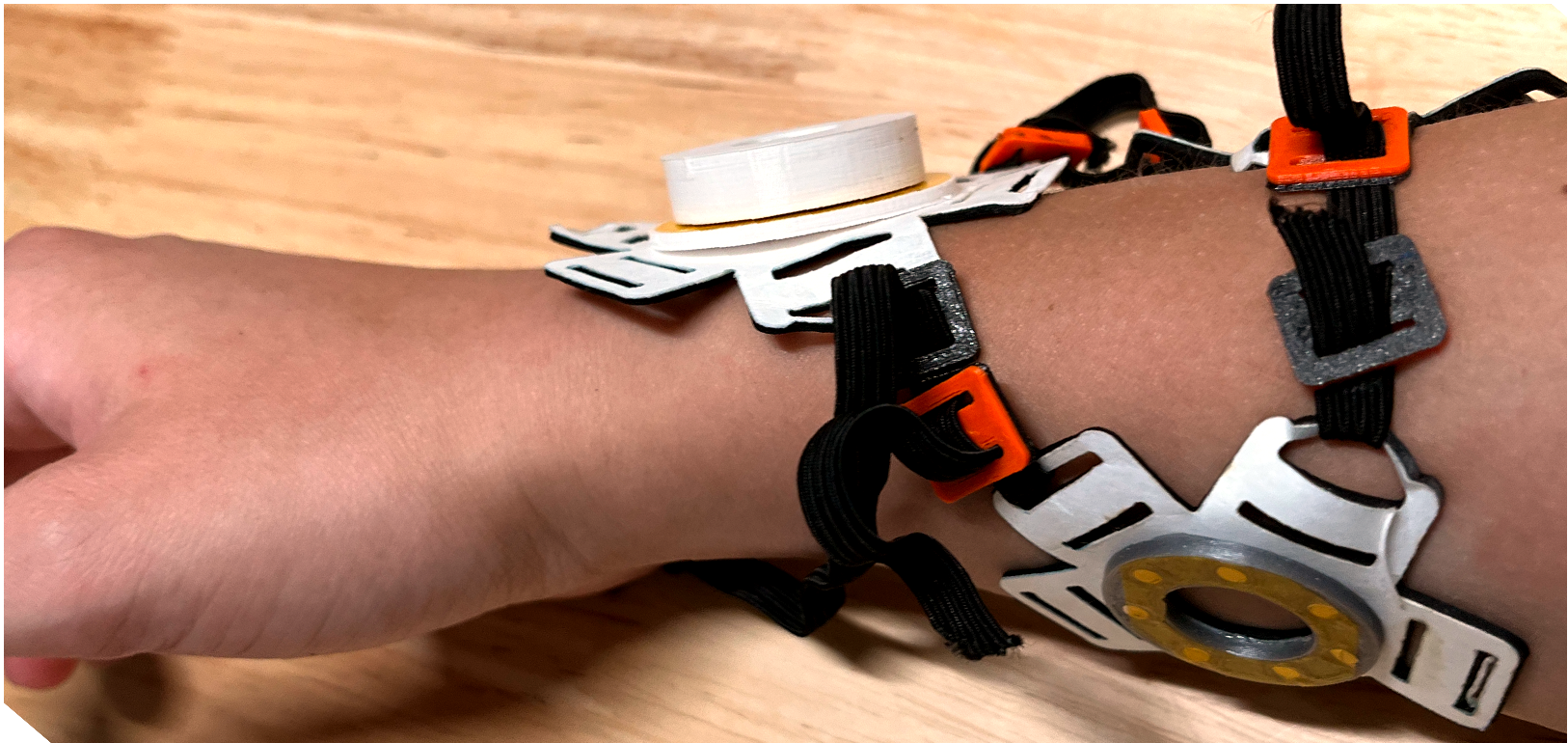}
            }
        }
    \end{tabularx}
    \caption{Current iteration of tile system}
    \label{fig:end-tile-system}
    \vspace{-4mm}
\end{figure}

A second strategy implements a gridded cuff that wraps around the limb (Fig. 3). Actuators screw into different positions along the grid. This design scheme lacks the modularity of the first strategy, but it is simpler to produce. 

\begin{figure}[h]
    \centering
    \hspace*{\fill}
    \subfloat[Back view of cuff\label{fig:back view of cuff}]{
        \includegraphics[width=0.35\columnwidth]{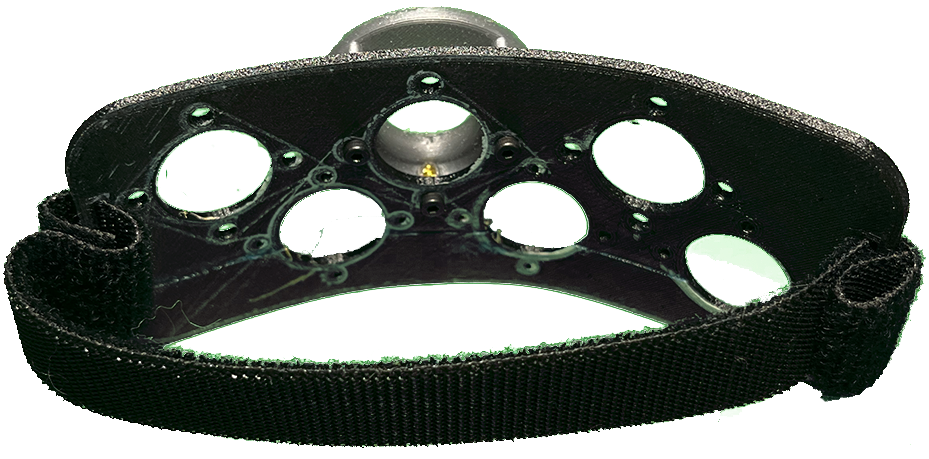}
    }
    \hspace*{\fill}
    \subfloat[Front view of cuff\label{fig:front view of cuff}]{
        \includegraphics[width=0.35\columnwidth]{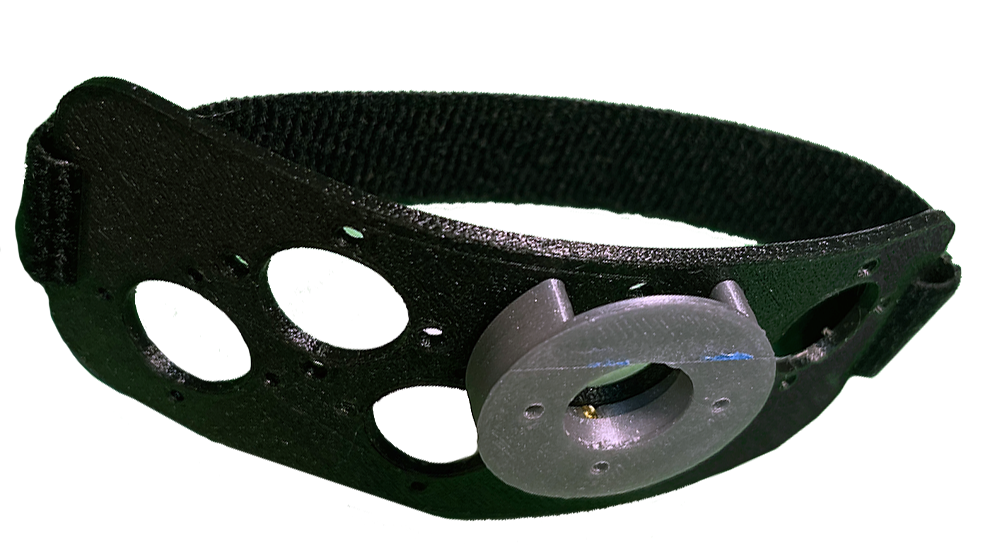}
    }
    \hspace*{\fill}
    \caption{The gridded cuff with an example actuator mounted}
    \label{fig:gridded cuff}
    \vspace{-2mm}
\end{figure}

In our third iteration, we have been exploring the use of a low cost vinyl cutter like \texttt{Cricut}\footnote{\url{https://cricut.com/en-us/}} along with \texttt{Tyvek} and EVA foam, instead of using laser cutter to improve accessibility of the system and reduce the overall production cost.

\section{Next Steps}
In the next phase of our project, we aim to conduct comprehensive human subject studies to assess the toolkit's practical applications. These studies will involve participants from varied backgrounds, focusing on collecting qualitative and quantitative data to evaluate the usability and effectiveness of the toolkit, i.e. address the key goals set forward in the design principles. The feedback will be crucial for identifying usability issues, understanding user experience, and pinpointing potential areas for improvement. Concurrently, we plan to undertake a detailed mechanical analysis of the toolkit components. This mechanical analysis will involve stress testing and simulations to evaluate durability, flexibility, and load capacity. The goal is to ensure the toolkit meets the necessary mechanical standards for robustness and longevity, reinforcing its reliability for extended use in research settings. 

As part of our continuous effort to refine and enhance the toolkit, we engage in various forums for expert feedback and collaboration. This engagement includes participating in relevant workshops\footnote{\url{https://uw-hapticexperincelab.github.io/HapticPlayground/}} and conferences to gather diverse perspectives. After refining the toolkit based on this feedback, we will make our toolkit open-source to enhance accessibility and encourage innovation in wearable haptic technology. Our findings will be published to support these efforts.


\end{document}